\documentclass[10pt]{iopart}
\usepackage{graphicx,epsfig}
\usepackage{bm}
\usepackage{iopams}
\usepackage{color,ulem}
\usepackage{enumitem}
\usepackage{graphicx}
\usepackage{appendix}
\usepackage{hyperref}
\usepackage{lipsum}
\usepackage{url}
\usepackage{mathrsfs}
\usepackage{amsbsy}
\usepackage{amstext}
\usepackage{amsthm}

\newcommand\eqref[1]{(\ref{#1})}

\usepackage{pdfsync}
\begin{document}

\title{Schouten-Codazzi Gravity}   
\author{ Roberto A. Sussman$^\dagger$ and Sebasti\'an N\'ajera$^\dagger$}
\address{$^\dagger$ Instituto de Ciencias Nucleares, Universidad Nacional Aut\'onoma de M\'exico (ICN-UNAM),
A. P. 70--543, 04510 M\'exico D. F., M\'exico.}
\eads{$^\ddagger$\mailto{sussman@nucleares.unam.mx}}
\date{\today}
\begin{abstract}
We propose a new phenomenological second order gravity theory to be denoted as ''Schouten-Codazzi' Gravity'' (SCG), as it is based on  Schouten and   Codazzi tensors. The theory is related, but is clearly distinct from Cotton Gravity. By assuming as source the energy momentum of General Relativity, we form a second order system with its geometric sector given by the sum of the Schouten tensor and a generic second order symmetric tensor complying with the following properties: (i) it must satisfy the Codazzi differential condition and (ii) it must be concomitant with the invariant characterization based on the algebraic structure of curvature tensors for specific spacetimes or classes of spacetimes.  We derive and briefly discuss the properties of SCG solutions for static spherical symmetry (vacuum and perfect fluid), FLRW models and spherical dust fluids.   While we do recognize that SCG is ``work in progress'' in an incipient stage that still requires significant theoretical development, we believe that the theory  provides valuable guidelines in the search for alternatives to General Relativity.\end{abstract}

\maketitle
\section{Introduction} \label{intro}

Detection of dark sector sources and understanding their fundamental properties is a major research topic in  contemporary theoretical physics. Unfortunately, progress on this task has remained elusive, thus motivating the search of gravity theories alternative to General Relativity.  Among the various alternatives, J. Harada \cite{harada2021emergence} proposed in 2021 a third order theory,  ``Cotton Gravity'', based on the third rank Cotton tensor \cite{cotton1899varietes,ellis2012relativistic}. Harada's theory can be derived from the conformal action \cite{mannheim1989exact} varying the connection and keeping the metric fixed. Harada showed that all solutions of General Relativity are also solutions of the new theory, derived a Schwarzschild-like exact solution and used \cite{harada2022cotton} the weak field limit to fit galactic rotation curves without assuming the existence of dark matter. 
\footnote{Harada proposed a separate theory denoted by ''Conformal Killing Gravity'' (CKG) \cite{Harada_2023a,Harada_2023b} (see also \cite{Mantica_2023}), which is a manifestly distinct from ``Cotton Gravity''.}.

However, Harada's Cotton Gravity theory has received little attention, since it  exhibits serious shortcomings \cite{clement2024cotton}: it cannot be applied to non-trivial conformally flat spacetimes  (such as FLRW models) because the Cotton tensor vanishes identically when the Weyl tensor vanishes \cite{stephani2009exact,garcia2004cotton}. It leads to underdetermination in highly symmetric spacetimes and presents ambiguities in identifying field sources (see \cite{clement2024cotton} and a response in \cite{response2024}).  An alternative formulation was found by Mantica and Molinari \cite{mantica2023codazzi}, based on the known fact that the field equations of Cotton Gravity result by demanding that the second order  Schouten tensor $S_{ab}$ and its associated source tensor ${\cal T}_{ab}$ (linearly related to the energy momentum tensor of GR) are both Codazzi tensors ({\it i.e.} satisfy the Codazzi differential condition).  

Mantica and Molinari also examined the mathematical properties of Codazzi tensors: their relation  with curvature tensors  \cite{mantica2012extended,mantica2012riemann,molinaricodazzi} and in \cite{mantica2023codazzi} their compatibility with different fluid sources and solutions of General Relativity. In another article these authors used their paramerization to examine and classify Friedman equations of various alternative theories \cite{Mantica:2023ssd}. However, as long as the Schouten tensor is subjected to the condition to be itself a Codazzi tensor, the parametrization of Mantica and Molinari is simply the usage of different intermediate variables for the same Cotton Gravity theory. 

 In the present article  we propose a second order phenomenological theory: ``Schouten-Codazzi Gravity'' (SCG),  that is related, but is clearly distinct from Cotton Gravity, since it follows from a modification of the roles of the Schouten and Codazzi tensors in the formulation of Cotton Gravity by Mantica and Molinari. We do not demand that  the Schouten tensor $S_{ab}$ and its source tensor ${\cal T}_{ab}$ to be both Codazzi tensors, as this would necessarily lead back to Cotton Gravity field equations. Instead, we form new second order field equations with the energy momentum of General Relativity and the geometric sector (left hand side) given by the Schouten tensor plus a generic symmetric tensor ${\cal F}_{ab}$ subjected to the following conditions: it must (i) encode the invariant symmetries that characterize the geometric sector of the field equations for specific spacetimes (or classes of spacetimes) and energy momentum tensors and (ii) satisfy the Codazzi differential condition. 
 
It is important to justify the specific build up of the generic tensor ${\cal F}_{ab}$. Condition (i) follows from the common algebraic structure of second order symmetric tensors in spacetime manifolds, which includes all Ricci-like tensors like its trace-free form, the Einstein and Schouten tensors \cite{hall1984classification,santos1995segre,bona1992intrinsic}. Since the generic ${\cal F}_{ab}$ must be prescribed  to determine the field equations, it must be functionally related {\it before solving the field equations}  to the eigenvectors and eigenvalues of the Ricci and Ricci-like tensors in an algebraic decomposition in orthonormal or suitable null tetrads. This decomposition is conventionally  encoded in a Segr\'e-Weiler and/or Pleba\~nski notation \cite{hall1976classification,cormack1979invariant} (see summary in Chapter 5 of  \cite{stephani2009exact}) and relates to all curvature and energy momentum tensors \cite{hall1986physical,hall1984determination,paiva1993limits,paiva1998limits,Zakhary1997ACS} (even in $f(R)$ theories \cite{afzal2024exact}). Condition (ii) allows to fully determine the functional relation between ${\cal F}_{ab}$ and eigenvalues of the Ricci or Ricci-like tensors through its compliance with the Codazzi differential condition. Besides this fact,  Codazzi tensors are Riemann and Ricci compatible, proven by Mantica and Molinari  in \cite{mantica2012extended,mantica2012riemann,molinaricodazzi} and also in their cited references. We also consider the results on the compatibility between Codazzi tensors and different forms of energy momentum tensors discussed in  \cite{mantica2023codazzi} and in Chapter 37 of \cite{stephani2009exact}.    
 
Following the steps outlined above, we probe the phenomenological SCG theory by obtaining exact solutions for three classes of spacetimes: static spherical symmetry (vacuum and perfect fluids), FLRW spacetimes and general spherical solutions with a dust source, with ${\cal F}_{ab}$ prescribed in terms of the eigenvalues of the Ricci or Einstein tensors before solving the field equations. These analytic solutions generalize known general relativistic solutions with same symmetries.  In the static vacuum case, we recover the Schwarzschild-like solution found by Harada \cite{harada2021emergence}, plus a solution that violates Birkhoff's theorem, but has unphysical features.  In static perfect fluid spacetimes the extra free parameters introduce significant modifications to the Tolman-Oppenheimer-Volkoff equilibrium equation. In the FLRW models SCG introduces an extra linear accelerating term that can be useful in observational Cosmology. We briefly comment on the physical properties of these solutions, but a deeper examination would require a separate article. 

While SCG does not require the computation of the Cotton tensor, we can still recover Cotton Gravity by applying the Codazzi differential condition to the second order field equations. Hence, all SCG solutions (including the ones we derived) are a restricted subset of solutions of Cotton Gravity, which being a third order theory contains many more solutions that are not solutions of the proposed theory. However, SCG solutions avoid the underdetermination seen in Cotton Gravity  \cite{clement2024cotton,response2024}.     

Evidently, the present article is a first step to explore the viability of the proposed SCG theory. We openly recognize that our proposal is still in an incipient stage, it is ``work in progress'  that requires much more theoretical development. There are still elements of ambiguity and potential underdetermination and we still need to show that its field equations  can be derived from a variational principle.  Since it is a second order theory, it avoids ``phantom'' solutions that plague hight order theories, also its field equations are much more tractable than those of Cotton Gravity or alternative theories of higher order derivatives. The shortcomings of Cotton Gravity  emerge from the properties of the Cotton tensor, but these can be avoided or (at least) significantly ameliorated because it does not require the computation of this tensor.  The theory we have presented is logically self-consistent and has a well defined correspondence limit with General Relativity (the vanishing of the prospective Codazzi tensor). Therefore, we believe that it merits further development.  

The section by section content of this paper is as follows: Section \ref{CGtoSCG} presents the field equations of SCG to highlight its  difference with Cotton Gravity.  Section \ref{invars} summarizes the invariant classification of second order symmetric tensors in 4-dimensional Lorentzian manifolds. Section \ref{codazzi} summarizes the properties of Codazzi tensors. Section \ref{efeE} discusses the problems of assuming an ``effective'' Einstein format for the field equations of SCG. Section \ref{sols} derives various exact solutions as 'test cases' to probe SCG. We summarize in Section \ref{properties} the geometric properties of the Codazzi and Schouten tensors. A final discussion and our conclusions are stated in Section \ref{final}, while \ref{AppA} and \ref{AppB} provide details on calculations done in Section \ref{sols}.

\section{From Cotton Gravity to Schouten-Codazzi Gravity}\label{CGtoSCG}

In its original paper Harada derived Cotton Gravity  directly from the Cotton tensor \cite{harada2021emergence}, but Mantica and Molinari have shown that this third order theory also emerges from the second order Schouten tensor through the Codazzi differential condition. It is necessary to understand this process to understand that the theory we are proposing in the present paper is distinct from Cotton Gravity.

\subsection{The Codazzi differential condition}\label{CDC}

Let ${\bf X}= X_{ab}{\bf e}_{\rm{\tiny{(A)}}}^a\otimes {\bf e}_{\rm{\tiny{(B)}}}^b$ be a second order symmetric tensor ($X_{ab}=X_{(ab)}$) in any basis of a smooth spacetime manifold ${\cal M}$. The Codazzi differential condition (or Codazzi ``deviation'')  can be defined by the map $\mathscr{F}: T_{{\cal M}} \to T_{{\cal M}} $ such that 
\begin{equation}\fl {\bf X}\to {\bf Y}=\mathscr{F}({\bf X})=Y_{abc}{\bf e}_{\rm{\tiny{(A)}}}^a\otimes {\bf e}_{\rm{\tiny{(B)}}}^b\otimes {\bf e}_{\rm{\tiny{(C)}}}^c,\qquad Y_{abc}=\nabla_b X_{ac}-\nabla_c X_{ab}.\label{codcond}\end{equation}
Notice that  ${\bf Y}$ is antisymmetric in $(ac)$ ({\it i.e} $Y_{abc}=-Y_{bac}$) :\,hence $g^{ab}Y_{abc}=0$. Also, $Y_{abc}$ satisfies $Y_{abc}+Y_{cab}+Y_{bca}=0$.  
   
\subsection{Cotton Gravity field equations}\label{CGeqs}

The Cotton tensor ${\bf C}$ results by applying the Codazzi differential condition \eqref{codcond} to the Schouten tensor ${\bf S}$:
\begin{eqnarray}\fl {\bf C}=\mathscr{F}({\bf S})=C_{abc}{\bf e}_{\rm{\tiny{(A)}}}^a\otimes {\bf e}_{\rm{\tiny{(B)}}}^b\otimes {\bf e}_{\rm{\tiny{(C)}}}^c,\quad  C_{abc}=\nabla_b S_{ac}-\nabla_c S_{ab},\label{CottonT},
 \end{eqnarray}
From this relation Mantica and Molinari \cite{mantica2023codazzi} obtained the field equations derived by Harada in the form:
\begin{equation} \mathscr{F}({\bf S}) =8\pi \mathscr{F}(\hat {\bf T}),\label{CGfeqs}\end{equation} 
where  ${\bf S}$ is the Schouten tensor and $\hat {\bf T}$ is  the source tensor, both given by
\begin{eqnarray}  {\bf S}=S_{ab}{\bf e}_{\rm{\tiny{(A)}}}^a\otimes {\bf e}_{\rm{\tiny{(B)}}}^b,\quad S_{ab}=R_{ab}-\frac16\,Rg_{ab},\label{SchoutenT}\\ \hat {\bf T}=\hat T_{ab}{\bf e}_{\rm{\tiny{(A)}}}^a\otimes {\bf e}_{\rm{\tiny{(B)}}}^b,\quad \hat T_{ab}=T_{ab}-\frac13\,Tg_{ab}, \label{hatT}\end{eqnarray}
where $R_{ab},\,\, T_{ab}$ are the components of the Ricci and energy momentum tensor and $R=R^a_a,\,\,T=T^a_a$ are their traces.  The tensor in the right hand side of \eqref{CGfeqs} 
\begin{equation}\fl  {\bf M}=\mathscr{F}(\hat {\bf T})=M_{abc}{\bf e}_{\rm{\tiny{(A)}}}^a\otimes {\bf e}_{\rm{\tiny{(B)}}}^b\otimes {\bf e}_{\rm{\tiny{(C)}}}^c,\quad  M_{abc}=\nabla_b \hat T_{ac}-\nabla_c \hat T_{ab},\label{eqM}\end{equation}
is a sort of generalized angular momentum \cite{gogberashvili2023general} that acts in Cotton Gravity field equations \eqref{CGfeqs} as a source for the Cotton tensor. As shown in \cite{response2024}, this tensor introduces ambiguity in identifying physical sources.  

\subsection{Schouten-Codazzi field equations}\label{SCGeqs}

To derive a second order theory, we consider the Schouten and source tensors ${\bf S}$ and ${\bf \hat T}$ as in \eqref{CGfeqs}, but without applying the Codazzi condition \eqref{codcond}, as this would lead back to Cotton Gravity:
\begin{equation}{\bf S}=8\pi \hat{\bf T}.\label{previous1} \end{equation}
Instead we consider modifying the geometric sector in the left hand side as, 
\begin{equation}{\bf S}+{\bf F}=8\pi \hat{\bf T}.\label{previous2}, \end{equation}
where ${\bf F}=F_{ab}{\bf e}_{\rm{\tiny{(A)}}}^a\otimes {\bf e}_{\rm{\tiny{(B)}}}^b$ is a  generic symmetric tensor whose components must be determined by solving  the differential Codazzi condition. Hence we will denote as the ``prospective Codazzi tensor'' (we discuss its characteristics in the following section):
 \begin{equation}\fl  \mathscr{F}({\bf F})=F_{abc}{\bf e}_{\rm{\tiny{(A)}}}^a\otimes {\bf e}_{\rm{\tiny{(B)}}}^b\otimes {\bf e}_{\rm{\tiny{(C)}}}^c=0, \qquad  F_{abc}=\nabla_a F_{bc}-\nabla_c F_{ba}=0.\label{codF1}\end{equation}
Since $\mathscr{F}({\bf F})$ is antisymmetric in the indices $(ac)$, we can associate to \eqref{codF1} a unique 2-form ${\bf f}$ and its closure condition $d{\bf f}=0$
 \begin{eqnarray}{\bf f}= f_b dx^b=\frac12 F_{abc}\,dx^a\wedge dx^c,=\frac12( \nabla_cF_{ba} -\nabla_aF_{bc})\,dx^a\wedge dx^c,\\
 d {\bf f}=\frac12 \nabla_d f_b dx^b=\nabla_a F_{bcd}\,dx^a\wedge dx^b\wedge dx^c=0,
 \label{codF2}\end{eqnarray}
leading to its six independent components to solve \eqref{codF1} and determine the field equations. .
 
 In terms of components, the field equations \eqref{previous2} read
  \begin{equation} R_{ab}-\frac16 Rg_{ab}+F_{ab}=8\pi \left(T_{ab}-\frac13Tg_{ab}\right),\label{FeqG}\end{equation}
 which can also be expressed in terms of the Einstein tensor as
  \begin{equation} G_{ab}-\frac13 Gg_{ab}+F_{ab}=8\pi \left(T_{ab}-\frac13Tg_{ab}\right),\label{SCGE}\end{equation}
 or in the Einstein ``effective'' forms as
  \begin{eqnarray} G_{ab}+F_{ab}-F\,g_{ab}=8\pi T_{ab},\label{SCGeff}\\
  R_{ab} +F_{ab}-\frac12 F g_{ab}=8\pi\left[T_{ab}-\frac12Tg_{ab}\right],\label{SCReff}\end{eqnarray}
 where $F=g^{cd}F_{cd}= F^a_a$. 
 
The contents of this section show that SCG is indeed a clearly different theory from Cotton Gravity (equations \eqref{CGfeqs}-\eqref{eqM}), though as shown in this Section both theories share also common features. Also, it is formally possible to pass the term $F^{ab}-F\,g^{ab}$ to the right hand side of \eqref{SCGeff} as a source term, with \eqref{SCGeff} or \eqref{SCReff} becoming Einstein equations with $F_{ab}-Fg_{ab}$ playing the role of an extra energy momentum tensor. We address  this issue in Section \ref{efeE} and the process to determine ${\cal F}_{ab}$ in Sections \ref{invars} and \ref{codazzi}. 

\section{The prospective Codazzi tensor and spacetime invariants}\label{invars}

For the proposed SCG theory to be viable the generic tensor ${\cal F}_{ab}$ (the prospective Codazzi tensor) must be prescribed as a pre-requisite before solving the field equation. As mentioned in the introduction, besides the requirement that this tensor complies with the Codazzi differential condition \eqref{codF1}-\eqref{codF2}, we will require that it relates to the invariant symmetries and scalars that characterize specific spacetime (or classes of spacetimes) and energy momentum tensors. 

\subsection{Concomitance with invariant properties.}\label{invarprops}

As a second order symmetric tensor in 4-dimensional Lorentzian manifolds $M$ the tensor ${\cal F}_{ab}$ shares the common algebraic structure of all such tensors, which admit invariant 2-dimensional subspaces that can be spacelike, timelike or null. These tensors can be described, examined  and classified through their their eigenvectors and eigenvalues determined from their matrix representation, either working with their characteristic polynomials \cite{santos1995segre,bona1992intrinsic}, or through their invariant 2-spaces in terms of suitable null tetrads obtained from any basis of the tangent space $T_p({\cal M})$ for all $p\in {\cal M}$ through the null rotation subgroup of the proper orthochronous Lorentz group \cite{hall1976classification,cormack1979invariant}.   

Eigenvectors and eigenvalues are labelled and classified in terms of  Segr\'e-Weiler or Pleba\~nski types.  For second order symmetric tensors  in 4-dimensions the Lorentz signature admits only 4 Segr\'e or Pleba\~nski types, which in the Segrr\'e symbol notation are (we follow the notation of  \cite{stephani2009exact}): [111,1], [211], [31] and [11,$z \bar z$]. where the digits inside the brackets denote the multiplicity of real eigenvalues, while  $z \bar z$ refers to a pair of complex conjugate eigenvalues of multiplicity one. Eigenvalues are all real in the first three types [111,1], [211], [31], admitting respectively  four, three, or two linearly independent eigenvectors, while the fourth type  [11,$z \bar z$] has two real and a pair of complex conjugate eigenvalues.  Degeneracy of eigenvalues is denoted by enclosing the corresponding digits in round brackets. For example [(11)1,1] and [(111),1] denotes two and three spacelike equal eigenvalues). Each Segr\'e type has a unique corresponding Plena\~nski type, but the latter explicitly provide the causal nature S,\,T,\,N  (spacelike, timelike, null). 

The eigenvector structure/classification described above is valid for all second order symmetric tensors in spacetime, including the Ricci, trace-free Ricci and Einstein (and Schouten) tensors, but the eigenvalues for a given Segr\'e or Pleba\~nski type are different for different tensors. In particular, these classification types rely on the eigenvectors and eigenvalue of the trace-free Ricci and energy momentum tensors \cite{hall1976classification,cormack1979invariant,hall1986physical}, while the classification Petrov types relies on conformal invariants of the Weyl  tensor \cite{hall1984determination} (see Chapters 5 and 6 of  \cite{stephani2009exact}). 

Since the generic tensor ${\cal F}_{ab}$ needs to be prescribed to determine the field equations for specific spacetimes, its form must be determined {\it before solving the field equation} through a functional relation with eigenvalues of the Ricci and Ricci-like tensors (all of which related through their common algebraic structure).  We describe  this process in Section \ref{codazzi} and apply it to derive solutions of SCG in Section \ref{sols}.  

Table 5.1 and equations (5.3a)-(5.3d) of \cite{stephani2009exact} summarizes the canonical form of the Ricci tensor  and Segr\'e types for the four allowed cases  (see Appendix A for computational details).
\begin{eqnarray}
\fl{\bf A1}:\quad  R_{ab}=\lambda_1 x_a x_b+\lambda_2 y_ay_b+\lambda_3 z_a z_b+\lambda_4 u_a u_b,\label{canRic1}\\
\fl \quad\quad  \hbox{Two real ortogonal eigenvectors. Segr\'e types:}\nonumber\\
\fl\quad\quad  [111,1], \,[11(1,1)],\,[(11)1,1],\,[(11)(1,1)],\,[1(11,1)],\,[(111),1],\,[111,1],\nonumber\\
\fl {\bf A2}:\quad R_{ab}=\lambda_1 x_a x_b+\lambda_2 y_ay_b+\lambda_3 k_{(a}l_{b)}+\lambda_4(k_ak_b-l_al_b),\label{canRic2}\\
\fl \quad\quad \hbox{No real eigenvectors. Segr\'e types:}\quad[11,z\bar z],\,[(11),z\bar z] \nonumber\\
\fl {\bf A3}:\quad R_{ab}=\lambda_1 x_a x_b+\lambda_2 y_ay_b-2\lambda_3 k_{(a}l_{b)}+k_ak_b,\label{canRic3}\\
\fl \quad\quad \hbox{One double null eigenvectors. Segr\'e types:} [11,2],\,[1(1,2)],\,[(11),2],\,[(11,2)],\nonumber\\
\fl {\bf B}:\quad  R_{ab}=\lambda_1 x_a x_b+\lambda_2 y_ay_b+\lambda_3 k_{(a}l_{b)}+k_{(a}x_{b)},\label{canRic4}\\
\fl\quad\quad \hbox{ Null invariant 2-spaces. Segr\'e types:} [1,3],\,[(1,3)],\nonumber
\end{eqnarray}
 where $\lambda_1,\,\lambda_2,\,\lambda_3,\,\lambda_4$ are the eigenvalues and $[{\bf x},\,{\bf y},\,{\bf z},\,{\bf u}]$ and 
 $[{\bf x},\,{\bf y},\,{\bf k},\,{\bf l}]$ are orthonormal and half null tetrads. Since the eigenvalues $\lambda_i,\,\,i=1,..,4$ are scalars we show in Sections \ref{codazzi} and \ref{sols} how to prescribe them as suitable functions to define the tensor ${\cal F}_{ab}$ before solving the field equations and in concomitance with \eqref{canRic1}-\eqref{canRic4}. 
 
 \subsection{Invariance of the energy momentum tensor}\label{EMT}

Energy momentum tensors share the common algebraic structure of all symmetric tensors in 4-dimensional spacetimes. While their invariant scalars can be specified beforehand, these scalars might be distinct from those of the curvature tensors (Ricci or Einstein) before solving the field equation.  In fact, the equality of invariant scalars of the curvature and energy momentum tensors occurs in the process of solving the   field equations.

The invariant algebraic reduction of the energy momentum tensor is discussed in detail by Hall et al in \cite{hall1986physical}. The authors examine various types of sources, including fluids, electromagnetic and radiation fields, as well as mixtures of them.  As a symmetric second order tensor in a Lorentzian manifold, the energy momentum tensor shares with Ricci and Ricci-like tensors the same algebraic structure and invariant canonical decomposition in terms of eigenvectors in the allowed Segr\'e types that we discussed before in \eqref{canRic1}-\eqref{canRic4}. However, the eigenvalues are different. In particular, the eigenvalues of  $T_{ab}$ are subjected to energy conditions: for any timelike vector field  $u^a$ we must have $T_{ab}u^au^b\geq 0$ and $T_{ab}u^b$ cannot be spacelike. These conditions forbid the Segr\'e types [3,1] and  [11,$z\bar z$] and restrict the eigenvalues of the remaining types. We summarize below the description provided in \cite{hall1986physical} for the cases of a fluid and electrovac energy momentum tensors (more sources were examined in  \cite{hall1986physical}).

\subsubsection{Fluids}  
Given a timelike 4-velocity field $u^a$ such that $u_au^a=-1$ and a projection tensor $h_{ab}=g_{ab}+u_au_b$, the energy momentum tensor of any fluid model is 
\begin{equation}T_{ab}=\rho u_au_b +p h_{ab}+\Pi_{ab}+2q_{(a}u_{b)},\end{equation}
where $\rho=u_au_bT^{ab}$ is the energy density, $p=(1/3)h_{ab}T^{ab}$ the isotropic pressure $\Pi_{ab}=\left[ h_a^{(c}h_b^{d)}-\
(1/3) h_{ab}h^{cd}\right]T_{cd}$ is the traceless spacelike anisotropic pressure tensor and $q_a=-T^{cd}u_dh_{ca}$ is the energy flux ($u_aq^a=u_a\Pi^{ab}=0$). For thermal systems, $\Pi_{ab}$ and $q_a$ are the shear viscosity and heat flux. We can identify the following particular cases:
\begin{itemize} 
\item Perfect fluid: $q_a=\Pi_{ab}=0$, then $u^a$ is a timelike eigenvector, with a Segr\'e type [(1,1,1),1] and eigenvalues $\rho,\,p$.
\item Fluid with anisotropic pressure without energy flux (viscosity without heat flux) $q_a=0$ with $\Pi_{ab}\ne 0$. Since $\Pi_{ab}$ is isometric to a spacelike 3-dimensional trace-less tensor, it can always be diagonalized as $\Pi^i_k=\hbox{diag}[P_1,\,P_2,-P_1-P_2]$. Hence, in this case the Segr\'e type is [111,1] with possible degeneracy of the spacelike eigenvalues $P_1,\,P_2$. 
\item Fluid with energy flux and isotropic pressure (heat flux without viscosity) $q_a\ne 0$ and $\Pi_{ab}=0$. In this case $u^a$ is  no longer an eigenvector, but together with $q^a$ it spans an invariant timelike 2-space, which rules out Segr\'e type [3,1]. Vectors in the orthogonal spacelike 2-space are eigenvectors with eigenvalue $p$. To examine this case, we define $q^a$ as parallel to one of the spacelike tetrads:  $z^a =zq^a$, leading to a modified tetrad $(l'^a,\,k'^a,x^a,y^a)$ with $\sqrt{2} l^a = z^a+u^a,\,\,  \sqrt{2} k^a = z^a-u^a$. The tensor $T_{ab}$ becomes
\begin{equation}\fl T_{ab}=\left[\frac12(\rho+p)+q\right]l'_al'_b +\left[\frac12(\rho+p)-q\right]k'_ak'_b+(p-\rho)l'_{(a}k'_{b)}+p(x_ax_b+y_ay_b),\end{equation}
where we can identify three subcases:
\begin{itemize}
\item $(p+\rho)^2<4q^2$, so that $p+\rho-2q<0<p+\rho+2q$. Rescaling $l',\,k'$ leads to complex eigenvectors eigenvectors $l^a\pm i\,k^a$ which is Segr\'e type [11,$z\bar z$] and is forbidden by energy conditions. 
\item $(p+\rho)^2=4q^2$. The dominant energy condition  and $q>0$ imply $p+\rho-2q=0$, hence it is Segr\'e type [2(11)] with eigenvectors $l^a,\,x^a,\,y^a$ and eigenvalues $p,\,p,\,(p-\rho)/2$.
\item $(p+\rho)^2>4q^2$. Dominante energy condition implies $p+\rho\pm 2q >0$. Rescaling $l.\,k$ leads to Segr\'e type [1(11),1] with spacelike eigenvectors $l+n,x,y$ and eigenvalues $-p+\epsilon,\,p-\epsilon,\,p,\,p$, where $\epsilon$ is given by
\begin{equation}\epsilon = \frac12\left(p+\rho\right)-\frac12\left[(p+\rho)^2-4q^2\right]^{1/2}>0.\end{equation}
\end{itemize}
\end{itemize} 

\subsubsection{Electrovac sources}. 

The energy momentum is 
\begin{equation}\fl T_{ab}=\nu l_a l_b\quad \hbox{null},\qquad T_{ab}=\mu(2l_{(a}k_{b)}-x_ax_b-y_ay_b)\quad \hbox{non-null,}\end{equation}
In the null case: $l^a$ is the principal null direction, Segr\'e type [(211)] with zero eigenvalue. The non-null case Segré type is  [(11)(1,1)] with eigenvalues  $\mu,\,-\mu$. Dominant energy conditions  implies $\nu>0$ and $\mu<0$. In the following section we discuss how the algebraic structure of the Ricci and and energy momentum tensors can guide the biuld up of the tensor ${\cal F}_{ab}$.  

\section{The Codazzi condition and properties of Codazzi tensors}\label{codazzi}
The algebraic structure of the Ricci tensor provides a rigorous guideline on how to determine the tensor ${\cal F}_{ab}$ before solving the field equations of the proposed theory \eqref{FeqG}-\eqref{SCReff}. Evidently, in the process of solving the field equations the eigenvalues of the tensors in the geometric  sector must be restricted by compatibility with eigenvalues of $T_{ab}$ for different sources, either with Einstein's equations or with the equations of the proposed theory  \eqref{FeqG}-\eqref{SCReff}. 

As an example to illustrate the determination of ${\cal F}_{ab}$, consider a spherically symmetric static perfect fluid source. The  Segr\'e type of the energy momentum tensor is [(111),1], which does not coincide with the Segr\'e type of the Ricci or trace-free Ricci tensor for this source before solving the field equations.  The symmetries of a static spherical metric before solving the field equations lead to a Ricci tensor  in an orthonormal base as \eqref{canRic1} with this specific form 
\begin{equation}R^a_b=R_1(r) x^a x_b+R_2(r) y^ay_b+R_2(r) z^a z_b+R_0(r) u^a u_b.\label{calF0}\end{equation}
with the eigenvalues $R_1(r),\,R_2(r),\,R_0(r)$ given by the three independent components of $R^a_b$ in this basis (see \ref{AppB}). This suggests a Segr\'e type [(11)1,1] for the Ricci or Ricci-like tensors before solving the field equations.  The form of $R^a_b$ in \eqref{calF0}suggests the following form for ${\cal F}_{ab}$  
\begin{equation}{\cal F}^a_b=F_1(r) x^a x_b+F_2(r) y^ay_b+F_2(r) z^a z_b+F_0(r) u^a u_b.\label{calF0bb}\end{equation}
with undetermined functions $F_1(r),\,F_2(r),\,F_0(r)$ that would enter in the field equations (for example in \eqref{FeqG} or \eqref{SCGE}) before solving them:
 \begin{equation} G_{ab}-\frac13 Gg_{ab}+F_{ab}=8\pi \left(T_{ab}-\frac13Tg_{ab}\right),\label{SCGEb}\end{equation}
with the Einstein and energy momentum tensors and their traces evaluated for the spherical static metric and the perfect fluid tensor. However, to solve the resulting field equations with ${\cal F}_{ab}$ given by \eqref{calF0bb} it is still necessary to determine the functions $F_1(r),\,F_2(r),\,F_0(r)$ in \eqref{calF0bb}. This can be done by imposing the Codazzi differential condition \eqref{codF1}-\eqref{codF2} on \eqref{calF0bb}, which leads to fully determined field equations. We follow this process to solve the field equations  in Section \ref{sols}.

\section{is Schouten-Codazzi Gravity just General Relativity with a modified source?}\label{efeE}

We address this question in this Section. The field equations of SCG can be expressed the Einstein ``effective'' forms \eqref{SCGeff}-\eqref{SCReff}
  \begin{eqnarray} G_{ab}+F_{ab}-F\,g_{ab}=8\pi T_{ab},\label{SCGeff2}\\
  R_{ab} +F_{ab}-\frac12 F g_{ab}=8\pi\left[T_{ab}-\frac12Tg_{ab}\right],\label{SCReff2}\end{eqnarray}
 where $F=g^{cd}F_{cd}= F^a_a$. This suggest transforming SCG into General Relativity with a second source by passing the terms with ${\cal F}_{ab}$ and ${\cal F}$ to the right hand side as sources given as a second energy momentum tensor, leading to the following Einstein's equations
 \begin{eqnarray} G_{ab}=8\pi \left(T_{ab}+T^{\hbox{\tiny{eff}}}_{ab}\right),\qquad T^{\hbox{\tiny{eff}}}_{ab}=\frac{-F_{ab}+F\,g_{ab}}{8\pi}\label{SCGeff3}\\
  R_{ab} =8\pi\left[T_{ab}-\frac12Tg_{ab}+T^{\hbox{\tiny{eff}}}_{ab}\right],\qquad T^{\hbox{\tiny{eff}}}=\frac{-F_{ab}+\frac12 F g_{ab}}{8\pi}.\label{SCReff3}\end{eqnarray}
This approach involves considering an added energy momentum tensor that is also a Codazzi tensor (if ${\cal F}_{ab}$ is Codazzi, so are the combinations ${\cal F}_{ab}-{\cal F}g_{ab}$ and ${\cal F}_{ab}-(1/2){\cal F}g_{ab}$).  

While this type of identification of these tensors as sources is formally possible and logically correct, it might be problematic.  First of all,  it is always formally possible in any alternative gravity metric theory to isolate the Einstein or Ricci tensors by``passing to the right hand side'' geometric terms as ''effective'' sources (there are examples in $f(R)$ theories). However, the main factor to judge this formal equivalence to General Relativity is to examine the possibilities for an appropriate physical interpretation of the new sources. 
 
 As we have shown in previous sections, keeping ${\cal F}_{ab}$ (or the terms with ${\cal F}_{ab}$ in \eqref{SCGeff2}-\eqref{SCReff2}) as part of the geometric sector (left hand side) allows us to consider the full availability of physical sources embodied in an unrestricted energy momentum tensor $T_{ab}$. As a contrast, considering the Codazzi tensors ${\cal F}_{ab}-{\cal F}g_{ab}$ and ${\cal F}_{ab}-(1/2){\cal F}g_{ab}$ as sources involves strong restrictions on these terms. Notice that both terms with ${\cal F}_{ab}$ in \eqref{SCGeff2} and \eqref{SCGeff3} are divergence-free free and are thus conserved, since $\nabla_bG^{ab}=\nabla_bT^{ab}=0$ hold.  However, this conservation involves a strong restriction on their functional forms:
 \begin{equation}\fl \nabla_b\left({\cal F}^{ab}-{\cal F}g^{ab}\right)=\nabla_b{\cal F}^{ab}-g^{ab} {\cal F}_{,b}=\frac{1}{\sqrt{-g}}\left[\sqrt{-g}{\cal F}^{ab}\right]_{,b}-\frac12g_{cd,b}{\cal F}^{cd}-g^{ab}{\cal F}_{,b},\end{equation}
which is further complicated by ${\cal F}_{ab}$ having to satisfy also the Codazzi differential condition \eqref{codF1}- \eqref{codF2}. 
As long as  ${\cal F}_{ab}$ (or ${\cal F}_{ab}-{\cal F}g_{ab}$) are in the left hand side, these restrictions do not affect the consistency of the field equations, they are satisfied as part of their solution for a given source $T_{ab}$. However, if ${\cal F}_{ab}$ (or ${\cal F}_{ab}-{\cal F}g_{ab}$) are taken as an energy momentum tensor describing a physical source, we have extremely restrictive conditions that are very likely leading to unphysical sources. 

The sources that can be described when demanding that the energy momentum tensor is a Codazzi tensor were examined by Mantica and Molinari in \cite{mantica2023codazzi}, showing that being a Codazzi tensor is a strong condition that severely limits the choice of  sources, even under assumptions of high symmetry. Among the restrictions, they found energy momentum tensors in static spacetimes that are proportional to flow tensors $\lambda(u^a\dot u^b+u^b\dot u^a)$ and $A u_au_b + Bg_{ab}$ with $\lambda,\,A,\,B$ scalars. The conditions for a perfect fluid  energy momentum tensor to be a Codazzi tensor impose strong restrictions:  a shear-free 4-velocity and homogeneous density in hypersurfaces orthogonal to the 4-velocity, which leads to a small subclass of exact solutions that includes the conformally flat Stephani universe. 

Mantica and Molinari used  in  \cite{Mantica:2023ssd} the Einstein effective equations \eqref{SCGeff3} to reproduce the field equations of various alternative gravity theories (Gauss-Bonet, teleparallel, $f(R)$ and others) in an FLRW geometry. However, they were successful in this task only because they remained in a FLRW geometry in which all energy momentum tensors are perfect fluids and these sppacetimes are conformally flat (like the Stephani universe). Reproducing the field equations of these theories using  \eqref{SCGeff3} would be either dauntingly hard or impossible in more general geometries.   Evidently, considering Einstein equations in the form \eqref{SCGeff3}-\eqref{SCReff3} might be formally possible, but it leads to extremely limiting and restrictive sources, while considering these terms as part of the geometric sector of SCG does not pose any restriction on the sources. 

\section{Exact solutions}\label{sols}
To probe the SCG theory we need to find exact solutions for specific geometries and sources following the prescription outlined  in the second paragraph of Section \ref{codazzi}. Our aim is to show that exact solutions can be consistently derived to compare with their analogues in General Relativity. We briefly comment on their properties, since a full discussion would require  a separate article. 

\subsection{Static vacuum spherical symmetry}\label{ssstvac}

The vacuum solution with spherical symmetry follows from the static  metric
\begin{equation}
ds^{2}=-e^{2\beta(r)}dt^{2}+e^{2\alpha(r)}dr^2+r^{2}(d\theta^{2}+\sin^{2}\theta d\phi^{2}),\label{staticSS}
\end{equation}
with $T_{ab}=0$. The energy momentum eigenvalues for vacuum $T_{ab}=0$ are trivially zero, therefore we need to rely on the most general eigenvalues of the Ricci or trace-free tensors of the metric \eqref{staticSS} to prescribe a form of ${\cal F}_{ab}$. This form given by \eqref{calF0} leading to \eqref{calF0bb}, which would be analogous for any other Ricci-like tensor, is the form to be used to determine and solve the field equation \eqref{SCGeff}. 

In fact, the same process takes place when solving Einstein's field equations for vacuum static spherical symmetry $R^a_b=0$, staring with $R^a_b$ given by its most general form \eqref{calF0} and leading to the Schwarzschild solution. Evidently, the vacuum field equations of SCG (\eqref{SCReff} with $T_{ab}=0$)
\begin{equation}R^a_b+{\cal F}^a_b-\frac12{\cal F}\delta^a_b=0,\label{SSvac}\end{equation}
will lead to different solutions from $R^a_b=0$. Applying the Codazzi differential condition \eqref{codF1}-\eqref{codF2} to   ${\cal F}_{ab}$ in \eqref{calF0bb} leads to two independent differential equations:
\begin{equation} F_0' = (F_1-F_0)\,\beta',\qquad F_2' = \frac{F_1-F_2}{r},\label{Cssst} \end{equation}
which contain no derivatives of $F_1$, suggesting to set it as $F_1=\lambda_0 =$ constant (as shown further ahead it will correspond to a cosmological constant), allowing to integrate \eqref{Cssst} as 
\begin{equation} F_2 = \lambda_0+\frac{\gamma_0}{r},\qquad F_0= \lambda_0 +\mu_0 e^{-\beta}.\label{C2C0},\end{equation}
where $\gamma_0,\,\mu_0$ are arbitrary real constants. Substitution of \eqref{C2C0} into the field equations \eqref{SSvac} leads after some algebra to the following system of three first order equations
\begin{eqnarray}
\alpha^\prime &=& \frac12(3\lambda_0 r+2\gamma_0)e^{2\alpha}+\frac{1-e^{2\alpha}}{2r},\label{eqar}\\
\beta^\prime &=& -\frac12\left[(3\lambda_0+\mu_0 e^{-\beta})r +2\gamma_0\right]\,e^{2\alpha}-\frac{1-e^{2\alpha}}{2r},\label{eqbr}\\
\alpha^\prime+\beta^\prime &=&-\frac12\mu_0 r e^{2\alpha-\beta}.\label{eqab}
\end{eqnarray}
which recovers the Schwarzschild solution of GR for $\lambda_0=\gamma_0=\mu_0=0$, so that $F_0=F_1=F_2=0$ with the Segr\'e type  [(1,\,1,1,1)]. For $\gamma_0=\mu_0=0$ but $\lambda_0\ne 0$, now identified as a cosmological constant, we recover the Schwarzschild-Kottler solution, also with Segr\'e type  [(1,\,1,1,1)]. However, for either one of $\gamma_0$ or $\mu_0$ nonzero we have solutions that are not General Relativity  solutions. To obtain the  general solution of \eqref{eqar}-\eqref{eqab} for nonzero $\lambda_0,\,\gamma_0,\,\mu_0$, we solve first $\alpha$ from \eqref{eqar} and substitute the result in \eqref{eqbr}, with the final result  
\begin{eqnarray}
\fl e^{2\alpha} &=& \left[1-\frac{2M_s}{r} +\gamma_0\,r+\lambda_0\,r^2\right]^{-1},\label{eqa}\\
\fl e^{2\beta} &=&\left[1-\frac{2M_s}{r} +\gamma_0\,r+\lambda_0\,r^2\right]\,\left[\mu_1+\frac{\mu_0}{2}\int{\frac{r^{5/2} \,dr}{\left[1-\frac{2M_s}{r} +\gamma_0\,r+\lambda_0\,r^2\right]^{3/2}}}\right]^2,\label{eqb}
\end{eqnarray}
where $2M_s$ (Schwarzschild mass) and $\mu_1$ are new integration constants.  Notice that the case $\mu_0\ne 0$ does not comply with  the Schwarzschild-like form $\alpha'+\beta'=0$ so that  $e^{\beta}=1/e^{\alpha}$.  If $\gamma_0\ne 0$ and $\mu_0=0$, then  $e^{\beta}=1/e^{\alpha}$ and we recover the Schwarzschild-like solution found by Harada \cite{harada2021emergence}.  

The quadrature in \eqref{eqb} has no analytic solution, but to examine the effect of this term multiplying $\mu_0$, we evaluate this integral for $\lambda_0=\gamma_0=\mu_1=0$, leading to the analytic form
\begin{eqnarray}
\fl e^{2\beta}=\frac{\mu_0^2}{16} \sqrt{B}\nonumber\\
\fl \times
 \left[
5 \mathit{M}_s r 
\left(B-\frac{4M_s}{r}\right)
+r^{2}+15 \sqrt{B}\, \mathit{M}_s^{2}
 \ln \, 
 \left(
 -\mathit{M}_s +r 
 \left(1+\sqrt{B}{r}+1\right)
\right)
\right]^{2} \end{eqnarray}
with $B= 1-2M_s/r$. This form of the metric coefficient $e^{2\beta}$ is clearly unphysical, as its asymptotic behavior as $M_s/r\ll 1$  at leading terms is $e^{2\beta}\sim r^4$. Therefore, we keep only the solutions with $\mu_0=0$. However, regardless of the unphysical nature of this particular solution,  \eqref{eqa} and \eqref{eqb} were obtained following a self-consistent procedure. 

\subsection{Static perfect fluid spherical symmetry}\label{ssstpf}  

We choose a more appropriate parametrization for the spherical static metric \eqref{staticSS} 
\begin{equation}
ds^{2}=-e^{2\Phi(r)}dt^{2}+\frac{dr^2}{1-\frac{\kappa M(r)}{r}}+r^{2}(d\theta^{2}+\sin^{2}\theta d\phi^{2}),\label{staticSS2}
\end{equation}
where $\kappa=8\pi G/c^2$ and we consider as source a perfect fluid
\begin{equation} T^{ab} = \rho u^a u^b + p h^{ab},\qquad u^a=e^{-\Phi}\delta^a_t.\label{PF}\end{equation}
The Segr\'e type of  the perfect fluid source is [(111),1], but as in the vacuum case we choose the  generic tensor ${\cal F}_{ab}$ as in \eqref{calF0bb}, of Segr\'e type [1(1,1),1] compatible with the form of $R^a_b$ for the metric \eqref{staticSS2} before solving the field equations. This leads through \eqref{codF1}-\eqref{codF2} to the same components of ${\cal F}_{ab}$   as in \eqref{C2C0}:
 \begin{equation} F_1=\lambda_0,\qquad F_0 = \lambda_0+\mu_0\,e^{-\Phi},\qquad F_2 = \lambda_0+\frac{\gamma_0}{r}.\label{C2C0b}\end{equation}
 Inserting \eqref{C2C0b} in the field equations \eqref{SCGeff2}: $G_{ab}-{\cal C}_{ab}+{\cal C} g_{ab}=8\pi T_{ab}$, together with the conservation law $\nabla_bT^{ab}=0$ leads to the following system of first order equations
\begin{eqnarray}
M^\prime &=& \left(4\pi \rho +\frac32 \lambda_0 \right)\,r^2 +\gamma_0 r,\label{c123a}\\
\Phi^\prime &=& \frac{8\pi p r^3 + 2M -\left[(3\lambda_0+\mu_0 e^{-\Phi})r+2\gamma_0\right]r^2 }{r(r-2M)},\label{c123b}\\
p^\prime &=& -(\rho+p)\Phi^\prime,\label{c123c}
\end{eqnarray}
which reduce to the  field equations of General Relativity for $\lambda_0=\gamma_0=\mu_0=0$. These are 3 equations for 4 variables $M,\,\Phi,\,\rho,\,p$, which becomes determined once an equation of state is chosen to relate $p$ and $\rho$ and  the free parameters that emerged as integration constants are fixed. The cosmological constant emerges by identifying $\lambda_0=(8\pi/3)\Lambda$, but nonzero $\gamma_0$ and $\mu_0$ represent degrees of freedom absent in General Relativity.

The two cases $\gamma_0\ne 0$ and $\mu_0\ne 0$ (with $\lambda_0=0$) represent interesting modifications of static solutions, which might be of astrophysical interest. If $\mu_0=0$, but $\gamma_0\ne 0$, both the mass integral \eqref{c123a} and the Oppenheimer-Volkoff equation are altered. In the case $\gamma_0=0$ with $\mu_0\ne 0$ the mass integral \eqref{c123a} is not affected, meaning that this case maintains the same relation between $\rho$ and $M$ as in General Relativity, though $\Phi$ and $p$ are affected. Also, the tensor ${\cal F}_{ab}$ has a dust form ${\cal F}_{ab}=\mu_0 e^{-\Phi}\delta_a^0\delta_b^0$.   

To examine the modifications from the cases when either $\gamma_0\ne 0$ or $\mu_0\ne 0$ hold, we consider their effect in rotation velocities $v(r)$ along timelike circular geodesics of the metric \eqref{staticSS2} in a toy model of a spherical galaxy. From the particle Lagrangian $-2{\cal L}=-\exp(2\Phi)\dot t^2+r_0^2\dot \phi^2$, we have  in General Relativity
\begin{equation} v^2(r) = r^2\,\dot \phi^2 = r\left[e^{2\Phi(r)}-1\right]\approx r\Phi^\prime\approx \frac{M(r)}{r}, \label{rotvel}\end{equation}
where we considered in the right hand side the weak field limit  conditions: $2M/r\ll 1,\,\,\Phi\ll 1$ and $p/\rho \sim  \sigma^2\ll c^2$ (with $\sigma^2$ the dispersion velocity and $\Phi$ the Newtonian potential).  Equation \eqref{rotvel} shows that in GR the only way to account for the observed increase of $v^2(r)$ in galactic rotation curves is an increase of $M$ (and thus of density $\rho$ from \eqref{c123a}), with the extra ``invisible'' mass also modifying $\Phi$. To see the effect of  the modifications from either $\gamma_0\ne 0$ or $\mu_0\ne 0$  on $v(r)$, we evaluate \eqref {rotvel} using the weak field expansion. We obtain the following weak field exressions 
\begin{eqnarray}\fl \gamma_0\ne 0,\,\,\, \mu_0=0:\quad v^2(r) \approx r\Phi^\prime \approx \frac{M(r)}{r}+\gamma_0 r,\quad M=4\pi\int{\rho r^2 dr}+\frac12\gamma_0r^2, \label{rotvelCG1},\\
\fl \gamma_0= 0,\,\,\, \mu_0\ne 0:\quad v^2(r) \approx r\Phi^\prime \approx \frac{M(r)}{r}+\mu_0 r^2,\quad M=4\pi\int{\rho r^2 dr},\label{rotvelCG2}\end{eqnarray}
revealing in both cases a correction to the Keplerian velocities in \eqref {rotvel}, corrections that are negligible at small scales close to the galactic center ($r\approx 0$) but becomes dominant at large values of $r$. In the case $\mu_0\ne 0,\,\,\gamma_0=0$ this modification does not affect $M(r)$, while in the case $\mu_0= 0,\,\,\gamma_0\ne 0$ the effect on $M$ depends on the sign of $\gamma_0$, for $\gamma_0>0$ it involves a larger mass increase together with an increase of rotation velocities than in a Newtonian regime, while $\gamma_0<0$ both the mass and the rotation velocities decrease. 

\subsection{FLRW models}\label{rw}

The metric of Friedman-Lema\^\i tre-Robertson-Walker (FLRW) models in rectangular coordinates is
\begin{equation} ds^2  -dt^2+a^2(t) \frac{dx^2+dy^2+dz^2}{\left[1+\frac{k}{4}(x^2+y^2+z^2)\right]^2}, \label{flrw}\end{equation} 
where $k=k_0 H_0^2|\Omega_0^{(k)}|,\,\,k_0=0,\pm\,1$ and the source is a perfect fluid $T^{ab}=\rho u^au^b+ph^{ab},\,\,u^a=\delta^a_t$. The Segr\'e type for the Ricci tensor and the energy momentum tensor is [1(1,1,1)], thus suggesting a generic tensor ${\cal F}_{ab}$ of the form
\begin{equation}{\cal F}^a_b=F(t) x^a x_b+F(t) y^ay_b+F(t) z^a z_b+F_0(t) u^a u_b.\label{calF0flrw},\end{equation}
so that applying the Codazzi differential condition \eqref{codF1}-\eqref{codF2} leads to
\begin{equation}F_0=\lambda_0,\quad F=\lambda_0+\frac{\gamma_0}{a}\end{equation}
with the field equations \eqref{SCGeff2} and the conservation law $\nabla_bT^{ab}=0$  taking the form  
\begin{eqnarray}
H^2=\frac{\dot a^2}{a^2} = \frac{8\pi}{3}\rho-\frac{k}{a^2}+\lambda_0+\frac{\gamma_0}{a},\qquad 
\dot\rho =-3(\rho+p)\frac{\dot a}{a},\label{flrw}
\end{eqnarray}
that can be integrated once an equation of state linking $p$ and $\rho$ is prescribed. Complementing  \eqref{flrw}, the Raychaudhuri equation is
\begin{equation}\frac{\ddot a}{a} =-qH^2 = -\frac{4\pi}{3}(\rho+3p)+\frac{\gamma_0}{2a}+\lambda_0. \label{q} \end{equation}
Also, setting $\rho=p=0$ in \eqref{flrw} and \eqref{q} we obtain (after some re-scalings)  the de Sitter ($\lambda_0>0$), anti de Sitter ($\lambda_0<0$) and Minkowski limits ($\lambda_0=0$). 

We can see from \eqref{flrw} and \eqref{q} that  $\lambda_0$ identifies with the cosmological constant, but the term $\gamma_0/a$ has no equivalent in General Relativity (it could only enter a general relativistic approach if introduced as an extra source). Notice that this term involves an accelerated expansion and appears added to the cosmological constant, so that cosmic acceleration would be produced by $\gamma_0/a+\lambda_0$ , which can be interpreted as a sort of correction to the acceleration predicted by the $\Lambda$CDM model (defined by $\gamma_0=0,\,\,\lambda_0 = (8\pi/3)\Lambda$ in \eqref{flrw}). The added term $\gamma_0/a$ can describe a contribution to the total cosmic acceleration that increases/decrease the $\Lambda$CDM acceleration for $\gamma_0$ positive/negative, with the numerical value of $\gamma_0$ adjusted to fit specific observations on the deviation from $\Lambda$CDM predictions.

\subsection{Spherical dust solution}\label{sphdust}

The exact solutions presented so far are characterized by second order symmetric but diagonal tensors. It is important to examine the case when curvature and/or energy momentum tensors are non-diagonal. A good example is spherically symmetric dust described by the following diagonal metric and energy momentum tensor  
\begin{eqnarray}ds^2=-dt^2+ A^2(t,r)dr^2+Y^2(t,r)[d\theta^2+\sin^2\theta d\phi^2],\label{ssdust}\\
T_{ab}=\rho u_au_b\label{Tabdust}\end{eqnarray}
This dust tensor is of Segr\'e type [(111),1] with a timelike eigenvector with nonzero eigenvalue and spacelike eigenvectors with eigenvalues equal to zero. The momentum conservation $h_{ca}\nabla_bT^{ab}=0$ implies $\dot u_a=u^b\nabla_b u_a=0$, hence the 4-velocity is geodesic with components $u^a=\delta^a_0$ in a comoving frame. Although $T_{ab}$ and $g_{ab}$ are diagonal, Einstein and all Ricci related tensors have nonzero off-diagonal components in the coordinate basis:
\begin{equation}G_{tt},\quad G_{rr},\quad G_{\phi\phi}=\sin^2\theta G_{\theta\theta},\quad G_{tr}=G_{rt},\end{equation}
while the components in the orthonormal basis $[x^a, y^a, z^a, u^a]$ are   
 \begin{eqnarray} G_{ab}=\left[\lambda_1 x_a x_b+\lambda_2 y_a y_b+\lambda_3 z_az_b +\lambda_0 u_au_n\right],\\
 \lambda_1 = \frac12\left[\left(G_{rr}+G_{tt}\right)-\sqrt{G_{rr}^2-2G_{rr}G_{tr}+4G_{tr}^2+G_{tt}^2}\right], \\
 \lambda_2=G_{\theta\theta},\quad \lambda_3=\sin^2\theta \lambda_2,\\
 \lambda_0 = \frac12\left[\left(G_{rr}+G_{tt}\right)+\sqrt{G_{rr}^2-2G_{rr}G_{tr}+4G_{tr}^2+G_{tt}^2}\right],
 \end{eqnarray}
 which clearly identifies a Segr\'e type [111,1] (would be [1(11),1] for the mixed tensor $G^a_b$) with the eigenvalues $\lambda_1,\,\lambda_0$ generating a timelike 2 space and two spacelike eigenvectors associated with $\lambda_2,\,\lambda_3$ (this Segr\'e type is analogous to that of the energy momentum tensor for a fluid with energy flux and isotropic pressure, discussed in Section \ref{EMT}). Solving Einstein's equation requires solving $G_{tt}=8\pi \rho$ but only after having solved the following 3 constraints
\begin{equation}
G_{rr} =0,\quad G_{\phi\phi}=\sin^2\theta G_{\theta\theta}=0,\quad  G_{tr}=G_{rt}=0,\label{constrGR}
\end{equation}
which then changes the Einstein tensor to a Segr\'e type [(111),1] compatible with the dust energy momentum $T_{ab}=\rho u_a u_b$. This process leads to the known Lema\^\i tre-Tolman-Bondi solution. 

For a spherical dust solution of SCG gravity for the metric \eqref{ssdust} and energy momentum tensor \eqref{Tabdust} it is necessary to follow the same process described above, but with the SCG field equations, that is with this replacement:
 \begin{equation} G_{ab} \quad \rightarrow\quad {\cal G}_{ab}= G_{ab}+{\cal F}_{ab}-{\cal F}g_{ab},\end{equation}
which would lead to solving ${\cal G}_{tt}=8\pi\rho$ after having solved the following constraints replacing \eqref{constrGR}
\begin{equation}{\cal G}_{rr} =0,\quad {\cal G}_{\phi\phi}=\sin^2\theta {\cal G}_{\theta\theta}=0,\quad  {\cal G}_{tr}={\cal G}_{rt}=0,\label{constrSCG}\end{equation}
whose solution requires prescribing the full set of nonzero components of  ${\cal F}_{ab}$, namely: ${\cal F}_{rr},\,{\cal F}_{\theta},\,{\cal F}_{tr},\,{\cal F}_{tt}$, as functions of $(t,r)$ that would then have to be determined by the Codazzi differential condition \eqref{codF1}-\eqref{codF2}. However, in general,  the Codazzi differential condition will involve derivatives of the metric and will have to be solved together with the field equations and the conservation equation $\nabla_b({\cal F}^{ab}-{\cal F} g^{ab})=0$ as a coupled system. We will examine this solution in a separate article. 

It is important to stress that for finding a consistent spherical dust solution for \eqref{ssdust}- \eqref{Tabdust} in SCG the constraints that must be solved are \eqref{constrSCG}, not \eqref{constrGR}. Evidently, if instead of dust we had considered another source, like (say) a fluid with energy flux that provides a better match with $G_{ab}$ it would be easier to identify the equations and constraints to solve. 

\section{Geometric properties of Codazzi and Schouten tensors and other options}\label{properties}

Since our proposed gravity theory is based on the Codazzi and Schouten tensors, it is necessary and useful to summarize the geometric properties of these tensors. One justification for the choice of these tensors is their link to Cotton Gravity, which is also based on them, but as explained in previous sections, we are working with them in the context of a second order theory that differs from their usage in Cotton Gravity. As far as we know, there is no theoretical ``silver bullet'', or an absolutely compelling reason for considering these two tensors (instead of other tensors)  as the theoretical basis to propose a gravity theory in the way we have described throughout the present article. However, as long as SCG is self-consistent and viable then further research and a proper consideration of their geometric properties  might provide useful arguments for a more solid justification for the choice of the Schouten and Codazzi tensors. 

\subsection{Codazzi tensors}

Codazzi tensors appear in the context of geometric compatibility conditions of immersed sub-manifolds. In particular, the theoretical basis of the 3+1 formalism of General Relativity involves the Gauss and Codazzi-Mainardi relations in the embedding of spacelike and timelike hypersurfaces in spacetime \cite{gourgoulhon20123+,gourgoulhon20073+}:
\begin{eqnarray}
\gamma^e_a\gamma^f_b \gamma^c_g\gamma^h_d \,{}^{(4)} R^g_{hab}={}^{(3)}R^c_{dab}+K^c_aK_{db}-K^c_b K_{ab},\label{gauss31}\\
\gamma^c_g n^h\gamma^e_a\gamma^f_b\, {}^4 R^g_{hef} = D_b K^c_a-D_a K^c_b,\label{codm31}
\end{eqnarray}
where $ {}^{(4)} R^g_{hef}$ is the 4-dimensional Riemann tensor of spacetime, while   ${}^{(3)}R^c_{dab},\,K^c_a,\,\gamma^a_b,\, n^c$ and $D$ are respectively the Riemann tensor, extrinsic curvature,  induced 3-dimensional metric, normal unit vector and the covariant derivative evaluated at the hypersurface.  The Codazzi-Mainardi relation-\eqref{codm31} relates the 4 dimensional curvature to a Codazzi condition \eqref{codcond} applied to the extrinsic curvature of the hypersurface. 

This embedding is a special case of  generic local isometric embedding of a manifold in another higher dimension manifold, including the case of interest in General Relativity: the embedding of a spacetime manifold $V_4$ in a n-dimensional flat space $E_n$ with $4<n<14$. There is an extensive literature on these embeddings \cite{gupta1975class,goenner1980local}, with an updated summary in chapter  37 of \cite{stephani2009exact} that discusses in detail the different embedding classes defined by the extra dimensions $p=n-4$ required for the embedding.  Following  \cite{stephani2009exact}, the  coordinates of $E_n$ and $V_4$ are respectively $y^A$ and $x^a$ with the parametric representation $y^A=y^a(x^a)$ and the metric of $V_4$ given by $ds_N^2 =y^A_{,a}y^B_{,b}dx^a dx^b$.  To examine the effect of $E_n$ on $V_4$ four vectors $y^A_{,a}$ tangent to $V_4$ and p unit vectors $n^{A(\alpha)}$ ($(\alpha) = 1..p$) orthogonal to $V_4$ are introduced. The covariant derivatives of  $y^A_{,a}\,\,n^{A(\alpha)}$ with respect to $x^a,\,g_{ab}$ are tensors in $V_4$. The condition $\nabla_c g_{ab}=0$ implies $\eta_{AB}y^a_{,c}\nabla_b y^B_{,a}=0$ so that  the covariant derivative  $\nabla_by^B_{,a}$ is orthogonal to $V_4$ and thus it can be expressed as linear combinations of the p normals $n^{A(\alpha)}$
\begin{equation}y^B_{,a;b} = \sum_\alpha e_{(\alpha)} \Omega^{(\alpha)}_{ab}\,n^{B(\alpha)},\qquad  \Omega^{(\alpha)}_{ab}= \Omega^{(\alpha)}_{ba},\label{yvsn}\\
\end{equation}
where the $e_{(\alpha)}$ are numeric constants and $\Omega^{(\alpha)}_{ab}$ are p symmetric tensors in $V_4$ that generalize to higher dimensions the role of the Codazzi differential condition \eqref{codcond} in defining the second fundamental form in hypersurfaces in spacetime (compare with \eqref{codm31}). Using \eqref{yvsn} and the relation $\eta_{AB}\,n^{A (\alpha)}Y^B_{,a}=0$ the derivative of the normals is
\begin{equation} n^{A(\alpha)}_{,a}=-\Omega^{(\alpha)}  g^{bc} y^A_{,c}+\sum_\alpha e_{(\beta)} t^{(\beta)(\alpha)}_a n^{A(\beta)}=0,\label{diffnA}\end{equation}
where $e_{(\beta)}$ are $p(p-1)/2$ constants and the vectors $t^{(\beta)(\alpha)}_a$  in $V_4$ are often called ``torsions''. The integrability conditions of \eqref{yvsn} and \eqref{diffnA} are generically known as the Gauss-Codazzi-Ricci relations that generalize \eqref{gauss31}-\eqref{codm31}:
\begin{eqnarray}
\fl R_{abcd}=\Sigma_\alpha e_{(\alpha)} \left[\Omega^{(\alpha)}_{ab} \Omega^{(\alpha)}_{cd}-\Omega^{(\alpha)}_{ad}\Omega^{(\alpha)}_{bc}\right],\qquad \hbox{(Gauss)},\label{gauss}\\
\fl \nabla_c\Omega^{(\alpha)}_{ab} - \nabla_b\Omega^{(\alpha)}_{ac}= \sum_\beta e_{(\beta)}\left(t^{(\beta)(\alpha)}_c\Omega^{(\beta)}_{ab}-t^{(\beta)(\alpha)}_b\Omega^{(\beta)}_{ac}\right),\qquad \hbox{(Codazzi)},\label{codemb}\\
\fl \nabla_b t^{(\beta)(\alpha)}_a-\nabla_a t^{(\beta)(\alpha)}_b=\sum_\gamma e_{\gamma}\left(t^{(\gamma)(\alpha)}_bt^{(\gamma)(\alpha)}_a-t^{(\gamma)(\beta)}_bt^{(\gamma)(\beta)}_a\right)\nonumber \\
\fl \qquad\qquad +g^{cd}\left(\Omega^{\beta)}_{cb}\Omega^{\alpha)}_{da}-\Omega^{\beta)}_{ca}\Omega^{\alpha)}_{db}\right),\qquad \hbox{(Ricci)},\label{ricciemb}
\end{eqnarray}
and their fulfillment determines the embedding of a given class p of generic spacetimes. However, as proven in \cite{gupta1975class,goenner1980local}, these conditions are not independent in some cases.  

It is interesting to examine conditions  \eqref{gauss}-\eqref{ricciemb} for the embedding classes $p=1$ and $p=2$. Both cases involve the Codazzi differential condition \eqref{codcond}, applied to a single symmetric tensor $\Omega_{ab}$ (class one) or to two tensors $\Omega_{ab},\,\Lambda_{ab}$ and a vector $t^c$ (class two)
\begin{eqnarray}
\fl \hbox{class one}\nonumber\\
\fl R_{abcd}=\pm \left(\Omega_{ac}\Omega_{bd} -\Omega_{ad}\Omega_{bc} \right)\quad \hbox{Gauss}\\
\fl \nabla_c \Omega_{ab}-\nabla_b \Omega_{ac} =0,\quad \hbox{Codazzi},\\
\fl \hbox{class two}\nonumber\\
\fl R_{abcd}=e_{(1)}\left(\Omega_{ac}\Omega_{bd} -\Omega_{ad}\Omega_{bc} \right)+e_{(2)}\left(\Lambda_{ac}\Lambda_{bd} -\Lambda_{ad}\Lambda_{bc} \right),\\
\fl \nabla_c \Omega_{ab}-\nabla_b \Omega_{ac}=e_{(2)}\left(t_c\Lambda_{ab}-t_b\Lambda_{ac}\right),\\
\fl \nabla_c \Lambda_{ab}-\nabla_b \Lambda_{ac}=-e_{(1)}\left(t_c\Omega_{ab}-t_b\Omega_{ac}\right),\\
\fl \nabla_b t_a-\nabla_a t_b=\Omega_{ac}\Lambda^c_b-\Lambda_{ac}\Omega^c_b,
\end{eqnarray}
These embedding classes are compatible with specific classes of spacetimes (see comprehensive discussion in \cite{stephani2009exact}), though there is little research on higher embedding classes, which limits their utility  to classify spacetimes.  

There is an extensive literature on Codazzi tensors outside of the realm of embeddings and sub-manifoleds, for example seminal historic research \cite{derdzinski1983codazzi,bourguignon2006codazzi} and more recent work updating it and providing new results  \cite{mantica2023codazzi, mantica2012extended, mantica2012riemann, molinaricodazzi,gruber2023parallel}. In particular, Mantica and Molinari \cite{mantica2012extended, mantica2012riemann} obtain new  appealing results on the  algebraic and differential relations of Codazzi tensors with curvature tensors. For example, when the Codazzi differential condition  \eqref{codcond}  (which these authors denote by the ``Codazzi deviation'') is applied as  $B_{ab}\to Y_{abc}=\nabla_a B_{bc}-\nabla_cB_{ba}$, where $B_{ab}$  is a generic symmetric tensor, it leads to its algebraic  Riemann compatibility:
\begin{equation}\nabla_a Y_{bcd}+\nabla_b Y_{cad} +\nabla _c Y_{abd}=B_{ae}R^e_{bcd}+B_{be} R^e_{abd}+B_{ae}R^e_{bcd}\end{equation} 
an identity that holds for $B'_{ab}  = B_{ab} +\chi \Omega_{ab}$ where $\chi$ is a scalar and  $\Omega_{ab}$ is a Codazzi tensor satisfying $Y_{abc}=0$. The closure of the 1-form \eqref{codF2} associated with the Codazzi differential condition $Y_{abc}$ leads to its fulfillment of the second Bianchi identity 
\begin{equation}\nabla_a Y_{bcd}+\nabla_b Y_{cad} +\nabla _c Y_{abd}=0\end{equation}
which (together with its contraction) provides the necessary and sufficient condition for Riemann and Ricci compatibility. 
\begin{eqnarray}B_{ae}R_{bcd}^e+B_{be}R_{cad}^e+B_{ce}R_{abd}^e=0,\\
\Rightarrow B_{ac}R_b^c-B_{bc}R_b^a=0.\end{eqnarray}
These are relations that will hold for the prospective Codazzi tensor ${\cal F}_{ab}$, while their proof is trivial for diagonal tensors, they hold in general. These properties  provide an appealing theoretical justification for identifying ${\cal F}_{ab}$ as a Codazzi tensor concomitant with the invariant properties of the Ricci tensor.  Also, applying the Codazzi differential condition \eqref{codcond} to the Ricci tensor leads to the divergence of the Riemann tensor
\begin{equation} \nabla_a R_{bc}-\nabla_b R_{ac}=-\nabla_d R_{abc}^d, \end{equation}
with a similar relation with the Weyl tensor
\begin{equation} {\cal F}_{abc}=\nabla_a {\cal F}_{bc}-\nabla_b {\cal F}_{ac} =-2 \nabla_d C_{abc}^d-\frac16 \left[g_{bc}\nabla_a R-g_{ac}\nabla_bR\right],\end{equation}
which leads to the Cotton tensor  if ${\cal F}$ is the Schouten tensor. 

\subsection{The Schouten tensor} 

The main reason for selecting the Schouten tensor \eqref{SchoutenT} in our proposal of SCG is its connection with Cotton Gravity. It is a curvature tensor closely related to the Ricci and trace-free Ricci tensors. It satisfies the same role as the trace-less Ricci tensor in the algebraic decomposition of the Riemann tensor in its Ricci and Weyl parts
\begin{equation}R^a_{bcd} = S_{ac}  g_{bd} +  S_{bd} g_{ac} - S_{ad} g_{bc} -S_{bc} g_{ad} +  C^a_{bcd},\label{SchW}
\end{equation}
There is interesting research on the fourth order tensor curvature formed with the Kulkarni-Numizu product of the Schouten tensor and the metric in \eqref{SchW} in terms of nonholonomic Riemannian manifolds in \cite{dragovic2003wagner,barrett2023schouten}, though these references use the term Schouten tensor for the fourth order tensor, as this tensor was specifically introduced by Schouten to examine nonholonomic Riemannian structures. Further important development on noholonomic curvature structures was due to Wagner, on the grounds that a vanishing fourth order Schouten tensor does not lead to flat nonholonomic structures, meaning  a path-independent parallel translation associated with the nonholonomic connection. This research did not influence our choice of the second order Schouten tensor in the proposed SCG theory. .   

\subsection{Other options}

Evidently, we could consider the added tensor ${\cal F}_{ab}$ in SCG to be determined by some other criterion besides being a Codazzi tensor. In particular, considering the relation between SCG to Cotton Gravity, it is worthwhile mentioning that Harada has proposed another third order gravity theory \cite{Harada_2023a} with similar arguments used in the derivation of Cotton Gravity, but now based on the conditions of conformal Killing tensors \cite{Harada_2023b,Mantica_2023}. Therefore, instead of demanding that the added tensor ${\cal F}_{ab}$  is a Codazzi tensor satisfying \eqref{codcond}, we could have demanded that it is a Killing or conformal Killing tensor \cite{cariglia2014killing,he2024spacetime}, or a Killing-Yano tensor \cite{dietz1981space,dietz1982space,batista2015integrability} thus leading to another second order theory with similar connection to that of SCG to Cotton Gravity. While this is certainly a valid possibility worth exploring in future research, we believe a Codazzi tensor choice for ${\cal F}_{ab}$ is for the time being a preferable and pragmatic choice, specially considering that Killing tensors involve hidden symmetries \cite{he2024spacetime}, while the choice of a conformal Killing  or KIlling-Yano tensor would involve dealing with too many  extra  free parameters.

\section{Final discussion and conclusions}\label{final}

We have presented a new alternative second order metric gravity theory that we have denoted by the term ``Schouten-Codazzi Gravity'' (SCG), as it is based on the Schouten tensor and a generic symmetric tensor that must be (i) concomitant with invariance scalars of curvature tensors and (ii) be a Codazzi tensor (satisfying the Codazzi differential condition). To  explain the features of the theory in reference to General Relativity, this generic tensor had to be defined in a precise and rigorous manner, which required reviewing known formalisms that provide an invariant classification of spacetimes and the properties of Codazzi tensors (Sections \ref{invars} and \ref{codazzi}). We also derived in Section \ref{sols} various exact solutions to probe the consistency of the theory and illustrate how it can work on specific test case spacetimes.  

While SCG is clearly distinct from Cotton Gravity (Section \ref{CGtoSCG}), the two theories have common elements: both theories rely on the Schouten and Codazzi tensors, but they work with them in completely different ways.  As shown by Mantica and Molinari \cite{mantica2023codazzi}, the Cotton tensor in the field equations of Cotton Gravity emerge from demanding that the Schouten tensor is a Codazzi tensor. Hence, in Cotton Gravity  the Schouten tensor is merely an intermediate second order tensor to compute the Cotton tensor. As a contrast,  SCG does not use the Schouten tensor to compute the Cotton tensor,  but to form the geometric sector of a new second order theory together with the generic symmetric tensor mentioned before. 

Not only SCG  by being second order is far more tractable than the third order Cotton Gravity, but SCG avoids its known shortcomings that have been noted in \cite{clement2024cotton,response2024}. As an example mentioned by these references, Cotton Gravity is underdetermined in highly symmetric spacetimes, such as vacuum solutions in spherical symmetry. Inserting the metric \eqref{staticSS} into the Cotton Gravity field equations \eqref{CGfeqs}, the constraint  $C_{abc}=0$ leads to a single third order equation for the two metric functions $\alpha$ and $\beta$:
\begin{eqnarray}
\fl \beta''' r^3+[(2\beta'-3\alpha')r +2] r^2\beta''+(3-2\alpha')r^2\beta'^2+[(2\alpha'^2-\alpha'')r^2-\alpha'r
\nonumber\\
\fl-2]r\beta' +(\alpha''-2\alpha'^2)r^2+1-e^{2\alpha}=0,\label{horrible}
\end{eqnarray}
where primes denote derivative with respect to $r$. Evidently, \eqref{horrible} leads to an under-determined problem with an infinite number of solutions. However, there are also problems of ambiguity of sources, since the solutions of $C_{abc}=0$ may not be vacuum because, the vanishing of the source term in Cotton Gravity field equations \eqref{CGfeqs} in \eqref{eqM} does not imply $T_{ab}=0$ \cite{response2024}. This ambiguity of sources does not occur in SCG, since the sources are exclusively defined by the same energy momentum tensor of General Relativity. 

As a contrast, the field equations of SCG for spherically symmetric vacuum \eqref{eqar}-\eqref{eqab} are fully determined. It is true that one of the particular solutions that were found (\eqref{eqb} with $\mu_0\ne 0$) is completely unphysical, but its derivation was consistent (many solutions found by alternative theories are also unphysical, this is just one example). However, the solutions \eqref{eqa}-\eqref{eqb} with $\mu_0=0$ are much less objectionable, since they fulfill Birkhoff's theorem and generalize  Schwarzschild and Schwarzschild-Kottler solutions of General Relativity.  SCG also avoids another problem of Cotton Gravity: its inability to handle non-trivial conformally flat spacetimes, such as FLRW models. As shown in Section \ref{rw}, solutions for these models  were found that introduce a linear term modifying the cosmic acceleration predicted by a $\Lambda$CDM model without (necessarily) introducing an extra source.  

We also addressed in Section \ref{efeE} an issue that might lead to misunderstanding. As other alternative theories, such as $f(R)$, it is formally possible to express the field equations of SCG  in an Einstein ``effective'' format, with the Einstein tensor isolated in the left hand side and passing to the right hand side all extra geometric terms as ``effective'' fluids added to the original energy momentum tensor $T_{ab}$. This way of creating Einstein equations is formally possible, but does not imply that the alternative theory is equivalent to General Relativity with a modified source. In fact, these ``effective'' fluids tend to be wholly inconsistent with physical assumptions, so they are mostly used as a tactic to solve field equations based on known techniques in handling Einstein's equations. In particular, the field  equations of SCG can be framed as Einstein's equations with a Codazzi tensor as a source added to $T_{ab}$. We argued in Section \ref{efeE} that an energy momentum tensor that is also a Codazzi tensor produces  ``effective'' fluids that are very restrictive and likely unphysical, while considering SCG as a alternative theory whose energy momentum $T_{ab}$.is the same as in General Relativity can be easily applied consistently to any source.  In section \ref{properties} we provided a summary of geometric properties o fht Codazzi and Schouten tensors, 

Finally, as mentioned before, we reiterate our full recognition that SCG is still in an incipient stage and requires more theoretical development and verification of its predictions in terms available empiric data. More general solutions need to be derived and examined. An urgent pending task is to find an appropriate Lagrangian and an action that lead to its field equations from a variational principle. Our aim in this article has been to show that SCG is self consistent, a necessary first step to undertake further research. 

\appendix

\section{Ricci eigenspaces}\label{AppA} 

The eigenvalues of the  trace-free Ricci tensor $\tilde R_{ab} = R_{ab}-(1/4) R g_{ab}$ and its associated  characteristic polynomial follow from solving 
\begin{equation} \tilde R^a _b v^b= \tilde \lambda v^a,\qquad \hbox{det}(\tilde R^a_b-\lambda \delta^a_b)=0\end{equation}
where the eigenvalue $\tilde\lambda$ relates to $\tilde\lambda$, the eigenvalue of the Ricci tensor, by    $\tilde\lambda = \lambda + (1/4)R$ (a similar relation holds for the eigenvalues of the Schouten tensor). In the Newman-Penrose tetrad  with  two null complex vectors $m^a,\,\bar m^a$ and two real spacelike vectors $n^a,\,l^a$,
\begin{equation}g_{ab}=-l_an_n-l_bn_a+m_a \bar m_b+m_b \bar m_a,  \end{equation}
we obtain the Ricci scalar invariants:
\begin{eqnarray}
\fl \Phi_{00} = \frac12 R^{a,b}l_a l_b,\quad \Phi_{01} = \frac14 R^{a,b}l_a m_b,\quad \Phi_{02} = \frac12 R^{a,b}m_a m_b,\quad \Phi_{12}= \frac12 R^{ab} m_a n_b,\nonumber\\ 
\fl \Phi_{11} = \frac12 R^{a,b}(l_a n_b+m_a \bar m_b),\quad  \Phi_{22} = \frac12 R^{a,b}n_a n_b,\quad \Lambda =\frac{1}{24}R=\frac{1}{24}R^c_c,\label{Phis}
\end{eqnarray} 
Segr\'e/Pleba\~nski types follow from the Ricci \eqref{Phis}  by means of specific algorithms (see comprehensive discussion in chapters 4, 5 and 9 of \cite{stephani2009exact}). In particular, these scalars, as well as the Conformal ones for the Petrov types, can be computed with the Tensor and Differential Geometry packages of Waterloo Maple Inc.\cite{maple}, based on algorithms derived by Letniowski and McLenaghan \cite{letniowski1988improved},  Zakhary and Carminati \cite{zakhary2004new} and Carminati and McLenaghan \cite{carminati1991algebraic}.

\section{Calculations on exact solutions}\label{AppB}

\subsection{Static spherical symmetry}

We evaluate the eigenvalues of the traceless Ricci tensor $\tilde R^a_b=R^a_b-\frac14 R\delta^a_b$ for the static metric 
\begin{equation}\tilde R^a_b v^b -\lambda v^a=0,\end{equation}
leads in general to three distinct eigenvalues
\begin{eqnarray}
\fl \lambda^0_0=-1-\frac{1}{2r^2}+\frac12 e^{-2\alpha}\left[-\beta'^2+\beta'\alpha'-\beta''-\frac{2(\beta'+\alpha')}{r}+\frac{1}{r^2}\right],\\
\fl \lambda^r_r=-1-\frac{1}{2r^2}+\frac12 e^{-2\alpha}\left[-\beta'^2+\beta'\alpha'-\beta''+\frac{2(\beta'+\alpha')}{r}+\frac{1}{r^2}\right],\\
\fl \lambda^\theta_\theta=\lambda^\phi_\phi=-1-\frac{1}{2r^2}+\frac12 e^{-2\alpha}\left[\beta'^2-\beta'\alpha'+\beta''-\frac{1}{r^2}\right],
\end{eqnarray}
which justifies the most general Segr\'e type [1(11),1] and the form of the prospective Codazzi tensor ${\cal F}_{ab}$. Notice that for Schwarzschild and Schwarzschild-like solutions $\alpha'+\beta'=0$ and $\lambda^0_0=\lambda^r_r$, leading to = [(111),1] and $\mu_0=0$ in \eqref{eqb}. 

The canonic NP tetrad the metric \eqref{staticSS} in the coordinate base $[dt,dr,d\theta,d\phi]$ is
\begin{eqnarray} \fl  l_a =\frac{e^{\beta} dt+e^{\alpha} dr}{\sqrt{2}},\quad  n_a =\frac{e^{\beta} dt-e^{\alpha} dr}{\sqrt{2}},\quad 
\frac{r\left(d\theta  + i\,\sin r d\phi\right)}{\sqrt{2}},\quad \frac{r\left(d\theta  - i\,\sin r d\phi\right)}{\sqrt{2}},\nonumber\\\end{eqnarray}
All Weyl scalars  vanish save $\psi_2$ corresponding to Petrov type D. The nonzero Ricci scalars \eqref{Phis} are
 \begin{eqnarray} \fl \Phi_{00}=\Phi_{22}=\frac{(\alpha'+\beta')e^{-2\alpha}}{2e},\quad \Phi_{11} = \frac{1+\left[r^2\left(\beta'^2-\beta'\alpha' +\beta''\right)-1\right]e^{-2\alpha}}{4r^2},
 \end{eqnarray}
 with $\Lambda$ the Ricci scalar. The possible Segr\'e types are determined by 4 conditions which are too cumbersome to display (the Maple worksheet is available by request), so we summarize the results:  type [ZZ(11)] is ruled out since $e^{-4\alpha}>0$. Type [(111),1] only occurs if $\alpha'+\beta'=0$, else for  $\alpha'+\beta'\ne 0$ the type is [1(11),1]. Hence, we consider this as most general Segr\'e type to form the tensor ${\cal F}^a_b$ in Sections \ref{ssstvac} and \ref{ssstpf}.


\section*{Acknowledgments}
SN acknowledges financial support from SEP–CONAHCYT postgraduate grants program.\\

\section*{References}


\bibliographystyle{unsrt}
\bibliography{CQG111361_arXiv}       

\end{document}